\documentclass[aps,prl,reprint,groupedaddress]{revtex4-1}
\usepackage{graphicx}
\begin{document}


\title{Type reproduction number for epidemic models
on heterogeneous networks
}


\author{Satoru Morita}
\email[]{morita.satoru@shizuoka.ac.jp}
\affiliation{Department of Mathematical and Systems Engineering, Shizuoka University, Hamamatsu, 432-8561, Japan}


\date{\today}

\begin{abstract}
Infection can spread easily on networks
with heterogeneous degree distribution.
Here, we considered targeted immunization on such networks, wherein
a fraction of individuals with the highest connectivity are immunized.
To quantify the effect of this targeted immunization approach on population immunity,
we proposed a method using the type reproduction number.
Consequently, we derived a precise and simple formula that can yield the immunization threshold,
which, to the best of our knowledge, is the first such result presented in the literature
\end{abstract}


\maketitle

In recent decades, advancements in the field of transportation have led to increased connectivity among people.
Owing  to this increased interaction, outbreaks of several new infectious diseases 
have occurred around the world, which are threatening the lives and health of  people.
In particular, these diseases spread over networks of individuals via contact between them.
In a similar manner, the spread of  computer viruses through the Internet has also
caused significant economic damages to affected individuals and entities. 
Therefore, there  is an urgent and important need to understand the mechanism
of these spreading phenomena in networks; moreover,
effective methods to control these infections are required.
A key issue for effective control of infections is to determine the groups of individuals on which preventive 
measures such as immunization should be focused .

In epidemiology, the basic reproduction number $\mathcal{R}_0$ has been
used to measure the transmission potential of infectious diseases\cite{anderson,diekmann}.
$\mathcal{R}_0$ represents the average number of secondary infections that a typical infection would directly
cause in a completely susceptible population. 
The standard method for calculating $\mathcal{R}_0$ involves determining 
the spectral radius $\rho(A)$ of the next-generation matrix $A$ for an infectious disease 
\cite{driessche,diekmann2010}.
When $\mathcal{R}_0>1$, the infection can spread in the host population;
in contrast, when $\mathcal{R}_0<1$, the infection will not spread.
Thus, $\mathcal{R}_0$ is a useful indicator of the effort 
required to eliminate an infection from the population.
If individuals in a host population are immunized at random, then the incidence of an infection will decline 
when the proportion of people with immunity exceeds $1-1/\mathcal{R}_0$,
which is referred to as herd immunity fraction \cite{anderson}.

However, the criterion by using $\mathcal{R}_0$ is based on the
assumption that the host population is homogeneous.
If the host population is structured, then  
the type reproduction number $\mathcal{T}$ is used in the place of $\mathcal{R}_0$ \cite{robert2003,heesterbeek2007,inaba2013}.
The type reproduction number represents
the average number of secondary infections in a subset that a typical infection in this subset would directly
cause in a completely susceptible population. 
If a vaccine is only applied administered to the subset  of the population,
the required fraction of vaccine coverage in this subset  can be given by 
$1-1/\mathcal{T}$, where $\mathcal{T}$ is the type reproduction number for the subset.
In the previous works \cite{shuai2013,lewis2019}, a simple method for deriving the type reproduction number
using the next-generation matrix has been proposed:  
if the next-generation matrix $A$ is decomposed into the target matrix $C$ of the terms subject to be immunized
and residual matrix $A-C$ of the terms not subject to be immunized, then we have 
\begin{equation}
\mathcal{T}_C=\rho(C(I-A+C)^{-1})
\label{TC}
\end{equation}
if $A$ is irreducible and $\rho(A-C)<1$ 
\cite{shuai2013,lewis2019}.

Considering the spread of infections in social networks,
an important property of networks that should not be
overlooked is its degree heterogeneity, 
where the degree $k$ is defined as the number of connections each node has with other nodes
\cite{newman2006,barabasi2016,RMP2002}.
It is well-known that the degree distribution often follows a power
law for large values of $k$:
\begin{equation}
P(k)\sim k^{-\gamma}.
\label{eq1}
\end{equation}
In this case, the network is called a scale-free network \cite{RMP2002,barabasi1999}.
For example, it has been reported that 
the networks of human sexual contact are scale-free
\cite{liljeros2001,schneeberger,ito2019a}.
On the contrary, some other studies on the subject have rejected this notion \cite{handcock,hamilton}.
While it is still being debated whether real sexual networks 
are strictly scale-free,
it is clear that they are highly heterogeneous;
this is because only a few individuals tend to have a large number of sexual partners,
while most individuals only have a few sexual partners.

In the popular susceptible-infected-susceptible (SIS) model in networks \cite{pastor2001a,pastor2001b,RMP2015}, 
the basic reproduction number is given as follows:
\begin{equation}
\mathcal{R}_{0}=\lambda \langle k^2 \rangle/\langle k \rangle,
\label{R0}
\end{equation}
where $\lambda$ represents the infection rate, which is defined later.
A similar formula for $\mathcal{R}_{0}$ has long been known in the field of epidemiology \cite{anderson,lloyd}.
If the degree distribution follows Eq.~(\ref{eq1}) and $\gamma\leq 3$,
then the second moment $\langle k^2 \rangle$ diverges in the large-size limit.
Thus, $R_0$ can diverge if $\lambda$ is finite.
Conversely, even if $\lambda$ is considerably small, 
the infection can become widespread.
While real social networks might not be strictly scale-free networks, typically, 
they have high $\langle k^2 \rangle$.

In this study, to develop efficient herd immunity, we considered the case wherein only 
a fraction of individuals in a population with the highest connectivity
($k\geq k_{\mbox{\tiny{max}}}$) are immunized; this is because 
it is expected that targeting individuals that act as hubs effectively reduces 
$\langle k^2 \rangle$.
Though this case has been analyzed in previous works \cite{pastor2002,RMP2015},
unlike those studies, herein, we quantify the effect of target immunization
by using the type reproduction number.
Furthermore, we also derive a new formula to calculate the immunization threshold.

To account for the effect of heterogeneity in the degree distribution of a population,
it is appropriate to consider the density $\rho_k(t)$ of infected
nodes within each degree class $k$.
Based on the previously proposed SIS model \cite{boguna2002,boguna2003},
the mean-field rate equation can be obtained as 
\begin{equation}
 \frac{d\rho_k(t)}{dt}=-\rho_k(t)+\lambda k
[1-\rho_k(t)]\Theta_k(t).
\label{eq2}
\end{equation}
In this equation, the first term on the right-hand side represents recovery, wherein 
the average duration of infection is set to one, while
the second term represents transmission, which is proportional to the combined product of 
infection rate ($\lambda$), density of susceptible nodes ($1-\rho_k(t)$),  
number of neighboring vertices ($k$), and probability that any
neighbor is infected ($\Theta_k(t)$).
In particular, the probability $\Theta_k(t)$ is the average of the probabilities that a connection from a node with degree $k$ exists to 
an infected node with degree $k'$ over all degrees:
\begin{equation}
 \Theta_k(t)=\sum_{k'}P(k'|k)\rho_{k'}(t),
\label{eq3}
\end{equation}
where $P(k'|k)$ represents the conditional probability that a node 
of degree $k$ is connected to a node of degree $k'$.
Assuming that there is no degree-degree correlation \cite{pastor2001a,pastor2001b}, 
$\Theta_k(t)$ could be considered independent of $k$, 
and thus, can be given as
\begin{equation}
 \Theta(t)=\frac{1}{\langle k \rangle}\sum_{k}kP(k)\rho_{k}(t).
\label{eq4}
\end{equation}
This is because, here 
\begin{equation}
P(k'|k)=k'P(k')/\langle k\rangle.
\label{enc}
\end{equation}

If the degree distribution has the maximum value $k_{\mbox{\tiny{max}}}$,
then the next-generation matrix of eq.~(\ref{eq2}) is as follows:
\begin{widetext}
\begin{equation}
A=\left(
 \begin{array}{cccc}
 \lambda P(1|1) & \lambda P(2|1) 
&\cdots &\lambda P(k_{\mbox{\tiny{max}}}|1) \\
 2\lambda P(1|2) & 2\lambda P(2|2) 
&\cdots &2\lambda P(k_{\mbox{\tiny{max}}}|2) \\
 \vdots & 	\vdots 
& \ddots       & \vdots \\
 k_{\mbox{\tiny{max}}} \lambda P(1|k_{\mbox{\tiny{max}}}) & k_{\mbox{\tiny{max}}}\lambda P(2|k_{\mbox{\tiny{max}}}) 
&\cdots &k_{\mbox{\tiny{max}}}\lambda P(k_{\mbox{\tiny{max}}}|k_{\mbox{\tiny{max}}}) \\
\end{array}
\right),
\label{NGM}
\end{equation}
\end{widetext}
where $A_{ij}$ represents the rate of infection for nodes of degree $i$ 
due to spread of the infection from infectious nodes of degree $j$.
The complete derivation of the matrix in Eq.~(\ref{NGM}) was performed using
the method proposed by Diekmann et al. \cite{diekmann2010};
we decomposed the Jacobian of Eq.~(\ref{eq2}) into
$T + \Sigma$, where $T_{ij}=iP(j|i)$ represents the transmission part, describing the production of new infections, 
and $\Sigma_{ij}=-\delta_{ij}$ is the transition part, describing changes in state, and computed $A=-T\Sigma^{-1}$.

If we target nodes with k larger than $k_t$,
the target matrix can be written as follows:
\begin{widetext}
\begin{equation}
C=\left(
 \begin{array}{cccc}
 0&0
&\cdots&0\\
 \vdots & 	\vdots 
&    & \vdots \\
 0&0
&\cdots&0\\ 
 k_t\lambda P(1|k_t) & k_t\lambda P(2|k_t) 
&\cdots &k_t\lambda P(k_{\mbox{\tiny{max}}}|k_t) \\
 \vdots & 	\vdots 
&        & \vdots \\
k_{\mbox{\tiny{max}}} \lambda P(1|k_{\mbox{\tiny{max}}}) 
& k_{\mbox{\tiny{max}}}\lambda P(2|k_{\mbox{\tiny{max}}}) 
&\cdots &k_{\mbox{\tiny{max}}}\lambda P(k_{\mbox{\tiny{max}}}|k_{\mbox{\tiny{max}}}) \\
\end{array}
\right).
\label{TM}
\end{equation}
\end{widetext}
Then, the type reproduction number 
$\mathcal{T}_{\geq k_t}$ is determined using  Eq.~(\ref{TC}).
In the absence of degree-degree correlation (i.e., Eq.~(\ref{enc})), by using
Eq.~(\ref{TC})),
the type reproduction number can be obtained as follows:
\begin{equation}
\mathcal{T}_{\geq k_t}=\frac{\frac{\lambda}{\langle k\rangle} 
\sum_{k=k_t}k^2 P(k)}{1-\frac{\lambda}{\langle k\rangle}
\sum_{k=1}^{k_t-1}k^2 P(k)},
\label{TRN}
\end{equation}
if the denominator is positive. 
If the denominator is negative, it means the divergence of $\mathcal{T}_{\geq k_t}$, i.e.,
the infection can survive even if all nodes of $k\geq k_t$ have immunity.
It is obvious from Eq.~(\ref{TRN}) that 
$\mathcal{T}_{\geq k_t}$ increases monotonically with respect to $k_t$.
Furthermore, if the entire population is targeted ($k_t$=1), 
the type reproduction number can be calculated as
\begin{equation}
\mathcal{T}_{\geq1}=\frac{\lambda }{\langle k\rangle}\sum_{k=1}k^2 P(k),
\end{equation}
which coincides with the formula for the basic reproduction number 
$\mathcal{R}_{0}$ given by Eq.~(\ref{R0}).
For a general case, it can be mathematically confirmed that 
$\mathcal{T}_{\geq k_t}>1\Leftrightarrow \mathcal{R}_0>1$ and 
$\mathcal{T}_{\geq k_t}<1\Leftrightarrow \mathcal{R}_0<1$
\cite{shuai2013,lewis2019}.

\begin{figure}
\begin{center}
\includegraphics[width=6.cm]{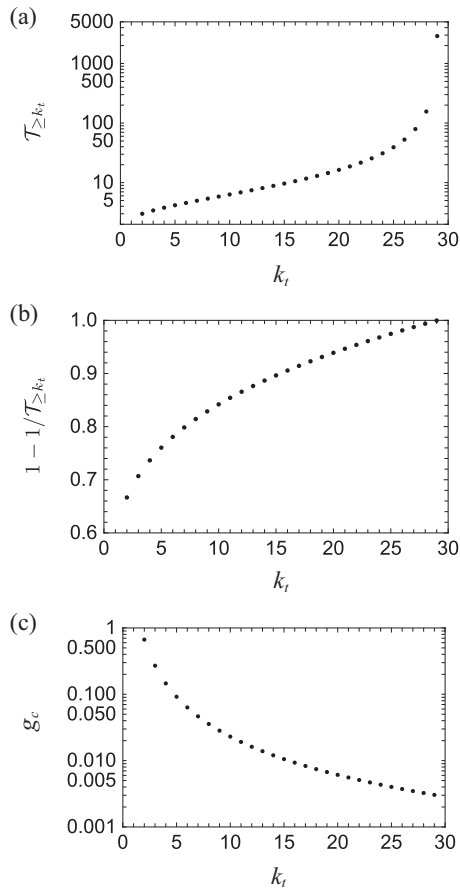}%
\caption{Characteristic curves for the type reproduction number:
(a) type reproduction number $\mathcal{T}_{\geq k_t}$ plotted as a function of $k_t$,
(b) plot for required fraction of immunized nodes with degree $k\geq k_t$,
and (c) plot of total amount of vaccine given by Eq.~(\ref{gc}).
Here, the degree distribution $P(k)\propto k^{-3}$ for 
$2\leq k \leq 10^4$ and the infection rate is set to
 $\lambda=0.22$ (such that $\mathcal{R}_{0}=3$). 
\label{f1}}
\end{center}
\end{figure}

We examine the characteristics of the type reproduction number
$\mathcal{T}_{\geq k_t}$, using the example shown in Fig.~1, where
the degree distribution $P(k)\propto k^{-3}$ for 
$k_{\mbox{\tiny{min}}}\leq k\leq k_{\mbox{\tiny{max}}}$
with $k_{\mbox{\tiny{min}}}=2$ and $k_{\mbox{\tiny{max}}}=10^4$.
It should be noted that $k_{\mbox{\tiny{max}}}$ is 
an artificially introduced cutoff; however,
a system with a finite size always has a similar cutoff.
The value of $\lambda$ is set such that $\mathcal{R}_{0}=3$;
consequently,
more than $1-1/\mathcal{R}_{0}=2/3$ of the total population would have to be randomly immunized
to prevent the spread of the infection.
Fig.~1(a) shows the dependency of $\mathcal{T}_{\geq k_t}$ on $k_t$;
in this case, because Eq.~(\ref{TRN}) is well-defined for $k_t\leq 29$,
the infection cannot be eradicated by immunizing only nodes with degrees $k>29$.
Thus, this critical value is based on the maximum value $k_t$ 
and satisfies:
\begin{equation}
\frac{\lambda}{\langle k\rangle}\sum_{k=1}^{k_t-1}k^2 P(k)<1.
\label{threshold}
\end{equation}

Then, the required fraction of the targeted 
nodes $k\geq k_t$ that need to be immunized can be obtained as follows:
\begin{equation}
1-\frac{1}{\mathcal{T}_{\geq k_t}}=\frac{\mathcal{R}_{0}-1}
{\frac{\lambda}{\langle k\rangle}\sum_{k=k_t}k^2 P(k)};
\label{TRN2}
\end{equation}
and tends to a value of one when $k_t$ approaches the critical value of 29 
as shown in Fig.~1(b).
In particular, this figure can be used to obtain the required value of $k_t$ based on public health constraints.
For example, if only 80\% of the target population can be vaccinated,
or the effective rate of vaccination is 80\%, then, to eradicate the infection, 
$k_t\leq 7$ 
because $1-\mathcal{T}_{\geq 8}>0.8$.

When all nodes with $k\geq k_t$ are immunized,
the proportion of the population that receives immunity from the infection is
$\sum_{k_t}P(k)$.
Because the total amount of vaccine is $\sum_{k_t}P(k)$ multiplied by 
$1-1/\mathcal{T}_{\geq k_t}$, it is calculated as
\begin{equation}
g_c=
(\mathcal{R}_{0}-1)\frac{\langle k\rangle}{\lambda}
\frac{\sum_{k=k_t}P(k)}
{\sum_{k=k_t}k^2 P(k)}.
\label{gc}
\end{equation}
It can be easily proved that $g_c$ is a decreasing function of $k_t$,
regardless of the degree distribution $P(k)$ (see also Fig.~1(c)).
Therefore, it was confirmed that the critical value of $k_t$ obtained via Eq.~(\ref{threshold}) or using its plot (such as in Fig.~1(b)) yields the optimal value for $k_t$.


In summary, we formulated an optimal immunization strategy, which 
is given by Eq.~(\ref{threshold}), based on
the degree and using the type reproduction number. 
The same immunization strategy has already been studied
by Pastor-Satorras and Vespignani
\cite{pastor2002,RMP2015}.
However, their reported formula for calculating
the immunization threshold
is different from the formula we obtained in this study, because
they focused on the number of links that disappeared
when the higher-degree nodes were removed, where
the fraction of disappearing links is given as follows:
\begin{equation}
p=\frac{\sum_{k=k_t}k P(k)}{\sum_{k=1} k P(k)}.
\label{pro}
\end{equation}
Then, they gave the immunization threshold as follows:
\begin{equation}
\frac{\langle k^2 \rangle_{g_c}}{\langle k \rangle_{g_c}}=
\frac{\sum_{k=1}^{k_t-1}k^2 P(k)}{\sum_{k=1}^{k_t-1}k P(k)}(1-p)+p<\frac{1}{\lambda},
\label{ef1}
\end{equation}
where $\langle \cdot \rangle_{g_c}$ represents the average of residual degrees after the links disappears.
In contrast, Eq.~(\ref{threshold}) can be rewritten as
\begin{equation}
\frac{\sum_{k=1}^{k_t-1}k^2 P(k)}{\sum_{k=1}^{k_t-1}k P(k)}(1-p)
<\frac{1}{\lambda}.
\label{ef2}
\end{equation}
The reason for this discrepancy between the previous work and current study is that, in the former case, 
it was assumed that links between nodes with $k<k_t$ can also disappear
with the probability given by Eq. (\ref{pro}); 
however, their assumption is not accurate because all links between nodes with $k\geq k_t$ 
must disappear too.
Thus, the authors of this previous study underestimated the critical value of $k_t$.
Accordingly, Eq.~(\ref{threshold}) provides a precise and simple formula
to calculate the immunization threshold.

Furthermore, while we considered the SIS model in our study, 
it is easy to extend our result to susceptible-infected-recovered (SIR) models for infections as well.
For the SIR model, the equation reported in Ref. \cite{bogua2003} can be used instead of Eq.~(\ref{eq3}), i.e.,
\begin{equation}
 \Theta_k(t)=\sum_{k'}\frac{k'-1}{k'}P(k'|k)\rho_{k'}(t).
\end{equation}
Consequently, Eq.~(\ref{threshold}) is replaced by 
\begin{equation}
\frac{\lambda}{\langle k\rangle}\sum_{k=1}^{k_t-1}(k^2-k) P(k)<1.
\label{threshold2}
\end{equation}


In conclusion, we showed that the type reproduction number is a considerably useful metric to
devise an optimal immunization strategy for a population.
It should be noted that 
the main result of this study, i.e., Eq.~(\ref{threshold}),
was obtained assuming no degree-degree correlation.
However, if degree-degree correlation is considered,
it is necessary to numerically calculate the type reproduction number
using the two matrices given by Eqs. (\ref{NGM}) and (\ref{TM}).
Lastly, the proposed method to calculate immunization threshold 
could also be used for various other extended epidemic models, such as in \cite{morita2016}.

\

\begin{acknowledgments}
This work was supported by the JSPS KAKENHI (no. 18K03453).
A part of this work was conducted at 
the Joint Usage / Research Center on Tropical Disease, Institute of Tropical Medicine, Nagasaki University (2019-Ippan-23), and at the Japan Science and Technology Agency Crest.
We would like thank Hiromu Ito for his valuable inputs for this study. 
\end{acknowledgments}

\bibliography{ref2020.bib}

\end{document}